# Appendix : 5

# Treatment of Reactive Routing Protocols using Second Chance based on Malicious behavior of nodes in MANETS

Mohammad Riyaz Belgaum[1], Safeeullah Soomro[1], Zainab Alansari[1]

[1]College of Computer Studies,
AMA International University,
Kingdom of Bahrain
[1]{bmdriyaz, s.soomro, zeinab}@amaiu.edu.bh

Muhammad Alam[2]
[2]Institute of Business and Management (IoBM),
Korangi Creek, Karachi, Pakistan.
malam@iobm.edu.pk

*Abstract*—Mobile nodes of various routing protocols in Mobile Ad hoc Networks follow different strategies in transmission and receiving of data. Security, packet delivery and routing overhead are important concerns for any protocol during designing them. The presence and absence of malicious nodes in the network affect a lot on the performance of the protocol. This research is mainly focused on the study of the threats, attacks and reasons for malicious behavior of nodes in the network for reactive routing protocols in MANETS. DSR and AODV are the two reactive routing protocols that were considered for the study to propose a second chance strategy to be given to the nodes considering the reason for malicious behavior to improve the packet delivery ratio and reduce the routing overhead in the network. A simulative study has been conducted using Ad hoc Simulator (ASIM) considering the DSR and AODV routing protocols in the presence of malicious nodes and in the absence of malicious nodes which showed that the packet delivery ratio is low and routing overhead is high in the absence of malicious nodes. The second chance strategy proposed considers the reasons for malicious behavior and helps the node to get reintegrated in the network to improve the packet delivery ratio and reduce the routing overhead.

I. INTRODUCTION

A set of mobile nodes that perform basic networking functions in a infrastructure less environment is said be a mobile ad hoc network (MANET). In the network each node helps another node in forwarding or routing packets to the destination.

Maintaining security is an important function of any of the routing protocol in each phase of the networking function [1]. Because of the non static topological behavior of the network and due to being the network open which allows the nodes to be added and deleted from the network anytime gives chance for the intruder nodes to disturb the normal routing process. And if there does not exist a common regulatory authority for authenticating and guaranteeing the nodes then a reliable transmission is not possible.

SECURITY RELATED GOALS AND CHALLENGES:

Security services are needed to make sure that the data is transferred over the network with reliability and also the keeping the resources of the system protected. To attain the objectives, the categorizations of security services are: availability, confidentiality, authentication, integrity and non-repudiation [2, 3].

• **AVAILABILITY:** Though the system is suffering from various problems like with bandwidth, connectivity but the availability service ensures that still the resources are available in a timely manner. Availability of a network can prove to harmful for example when the packets are dropped and by resource depletion attacks.

• **CONFIDENTIALITY:** The information prevailing in the network is not to be shared among all unauthorized nodes and this is achieved by Confidentiality. In order to achieve Confidentiality many encryption techniques can be used to make only the authorized nodes can share the transmission of information and the private and public keys.

• **AUTHENTICITY:** To prove a node as a legitimate user the network service used is Authenticity. The absence of this service can make any node in the network impersonate any node, and then having a total control capture and control over the complete network.

• **INTEGRITY:** The data which is been transmitted in the network can be modified either wontedly or sometimes unwontedly. The Integrity network service ensures that the information which is been transmitted is not modified.

• **NON-REPUDIATION:** This service guarantees that the message transmission has been done between the two parties

and it cannot be denied. Also using this service it helps in detecting and isolating of compromised nodes in the network.

Communicating through the network in safe and secure way has been a challenging task because of

- Not being a stable infrastructure.
- The links in the network are prone to break and not secure.
- Scarcity or overload on the system resources
- The network topology being dynamic.

A study of the various on demand routing protocols in MANETS has been done to analyze the threats and types of attacks in the routing protocols along with the reasons for the nodes to be malicious. The proposed approach will be used to reintegrate the node back into network which were previously as malicious.

## II. REVIEW OF LITERATURE

To have a secure transmission various secure routing protocols have been designed for MANETS [4, 5, 6, 7, 8]. The features and functionality of the routing protocols are studied from [9]. Based on their type and the functions they perform, the protocols are categorized as follows.

| Ad hoc Routing Protocols | | |
|---|---|---|
| Category | Type | Name of protocol |
| Flat Routing | Proactive | OLSR |
| | | DSDV |
| | | WRP |
| | Reactive | DSR |
| | | AODV |
| | | TORA |
| | | ABR |
| | | SSR |
| Hierarchical Routing | | HSR |
| | | CGSR |
| | | ZRP |
| | | LANMAR |
| Geographic Position assisted Routing | | GEOCAST |
| | | LAR |
| | | DREAM |
| | | GPSR |

### A. Reactive Routing Protocols:

*i. Dynamic Source Routing Protocol:*

The Dynamic Source Routing protocol a reactive routing protocol has two phases namely route discovery and route maintenance [10]. Initially the routes are discovered for transmitting the packets between source node and the target node. A route cache is maintained having the information of the recently used routes. In the route maintenance phase it ensures that the packet is safely transmitted to the target as there is a chance of routes being broken due to the dynamic nature of the topology. More over the researcher simulated the protocol using NS-2 to focus on the performance of this protocol using the metrics like packet delivery ratio in the presence of malicious nodes and in the absence of the malicious nodes. It was found that the packet delivery ratio in the presence of malicious nodes is more than in absence of malicious nodes.

*ii. Ad hoc On Demand Distance Vector Protocol:*

The AODV protocol explains its functionality [11]. It is stated that the features of both DSR and DSDV are combined. The author explains the working of AODV protocol along with the Black hole attack and its effect on the protocol. Finally a Counter Algorithm is proposed to prevent the Black hole attack on make the AODV Protocol secure. Metrics like Packet Delivery Ratio and Average End-to-End Delay was used to measure the performance of the protocol. A method to identify the malicious node was explained in order to avoid forwarding of the information to the malicious node in the routing table. The solution given did not impose any overhead on the nodes in the network.

*iii. Temporally Ordered Routing Algorithm:*

The Temporally ordered Routing Algorithm works using the link reversal concept and this algorithm ensures that it results in no loops [12]. The functionality of this protocol works in three phases as follows: (a) Phase – I creates the route initially (b) Phase – II concentrates on the maintenance of the routes and (c) Phase – III deletes the routes that are proved to be invalid. TORA protocol was compared with other routing protocols like DSDV, OLSR with the metrics like Control Overhead, Packet Delivery Ratio, End to End Delay and throughput that it is better in performance with varying number of nodes using NS-2 simulator.

*iv. Associativity Based Routing:*

The Associativity Based Routing protocol does not have loops, free of deadlock and no duplicate packets [13]. It focuses on route longevity. As there are very few broken communication links and less need for reconstruction of the routes the overhead involved is less. An improved version of ABR was to optimize the bandwidth and demand to reduce the overhead based on the position information was proposed. It was concluded that the path setup time was long for the routes which gave a scope for the future research to improve the ABR Protocol.

*v. Signal Stability-based Adaptive Routing Protocol:*

The working of SSR routing protocol states that the large routing tables are not required for routing [14]. The network will not be congested with the control messages but a type of denial of service attack is a threat to this protocol. The Signal Stability Table is maintained that has information of signal's strength of all nearby nodes. The protocol was simulated in OmNet and a metric known as CPU usage was considered to measure the performance. It proved that in the presence of malicious node the usage of CPU was more than in the absence of malicious nodes.

### B. TYPES OF ATTACKS:

Attacks are broadly classified into two categories as Passive attacks and Active attacks based on whether the operation in the network is disturbed or not [15].

Passive attacks allow the unauthorized node to snoop the data without altering it. The main target of the intruder node is to steal any information that is being transmitted which in turn makes the confidentiality to fail. These types of attacks are very difficult to detect but can be prevented by using powerful encryption mechanisms.

Active attacks modify or even delete the data which is being transmitted through the network. The affect of these types of attacks is that it brings down the performance of the network. The following is a brief description of various active attacks.

**Sybil Attack:** In this type of attack the intruder node pretends to have multiple identities and modifies and alters the information by directly communicating with the legitimate nodes [16].

**Resource Consumption Attack:** A malicious node in this attack always consumes the system resources like the bandwidth, battery power, memory, access points etc. A form of replication of packets makes the nodes to consume bandwidth and also the battery power.

**Black Hole attack:** The intruder node in this attack announces that to reach any other node, it has the nearest path in the network. Whenever the intruder node receives a request to reach the destination, this node replies saying that it has the shortest route. If this reply message reaches to the node which initiated the request earlier than the actual node then the route is established and hence the intruder node has found the place between two actual communicating nodes causing denial of service behaving as man in the middle.

**Worm Hole attack:** The attacker in the worm hole attack broadcasts to the other nodes as a neighbor after hearing transmission of packet from some node. The malicious node now forwards the packet to another location creating a worm hole. These types of attacks prove to be very dangerous [17].

### III. METHODOLOGY

The main aim of this research is to identify the various characteristics and approaches of the reactive routing protocols and study the attacks and the reasons for malicious behavior of nodes in reactive routing protocols in order to effectively propose a general framework for the same. The study aims to consider the reactive routing protocols with different types of attacks on them.

The study addresses the following questions:

1. Reasons for the malicious behaviour of the nodes.
2. The effect of having malicious nodes in the network and eliminating malicious nodes in the network.

After the reasons for malicious behavior of the nodes are studied, the process of treating such nodes following a common strategy can be developed which may help in maximizing the throughput and minimizing the routing overhead is answered.

The study mainly uses the following reactive routing protocols

- DSR (Dynamic Source Routing Protocol)
- AODV (Ad hoc On Demand Distance Vector Routing Protocol)

**Factors Considered are**

- Throughput
- Average delay
- Routing overhead

A simulative and analytical research methodology has been adopted in conducting the study. The various routing protocols have their own methodologies to send the information to the destination. This research considers the reasons for eliminating the nodes from the path due to the malicious behavior. The plan is to propose a strategy which will help in maximizing the throughput and minimizing the routing overhead on the routing protocols and thus help in the selection of the most optimal routing path for any protocol to send the information to the destination.

The research is organized as:

- Study the reactive routing protocols
- Analyze the reasons for malicious behavior of the nodes
- Propose a strategy to handle malicious behaviour

### IV. DISCUSSION

The prominent objective of any ad hoc routing protocol is to have a safe and secure transmission which has become a challenging task because of (1) lack of security during communication, (2) changes in the network structure or topology, (3) resource constraints (e.g. battery power, bandwidth, memory, and CPU processing capacity) and (4) movement of node that makes the network topology dynamic. Most of the traditional routing protocols design could not take into account the aspect of security.

The reasons for malicious behavior of the nodes are studied and protection can be provided to keep secured and to handle these types of misbehavior like not forwarding the data or messages, Though there is a secure and safe route traffic is deviated, Though there is no error packets are rerouted, Not observing the errors and lack of sending error messages, or vice versa, frequently updating the routes unusually.

One of the malicious behaviors like not forwarding the data or messages is considered in this research. Again there could be many reasons for a node not to forward the messages or data like Nodes failure, link and network failures, transmission energy and power failures.

Simulations are implemented using Ad hoc Simulator (ASIM), a simulator which is developed to simulate protocols in MANETS. ASIM is developed for wireless ad hoc networks. The Physical Terrain Dimensions considered is 500m X 550m. The range of communication is from 100m to 200m. Considering the number of nodes from 30 to 50 to be varying due to dynamic topology in a network, the speed is uniformly distributed between 0 and 20 m/s to offer a range of users that are in a fixed location or in motion. In this simulation, nodes move according to the random waypoint mobility model. The data communication pattern in this study uses multiple source destination pairs, each sending a constant Bit Rate (CBR) flow of 4 data packets per second. The simulation time is chosen to be long enough to potentially roam the whole area. The Application data packet size is 512 bytes. The factors varied are the total number of nodes in the network, the transmission time, Communication Range and the different network topologies.

Metrics like Packet Delivery Ratio and Routing Overhead are considered both in presence of malicious nodes and in absence of malicious nodes to compare the performance of reactive the routing protocols selected in this research.

Packet Delivery Ratio = Total no. of packets delivered / Total no. of packets sent.

Overhead = Total no. of routing messages transmitted / Total no. of routing messages received

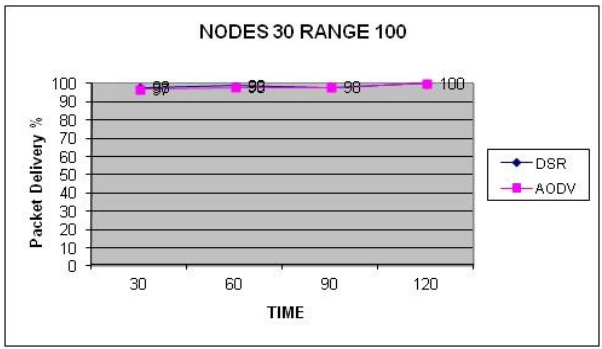

Fig 1: Packet Delivery Ratio in the presence of malicious nodes.

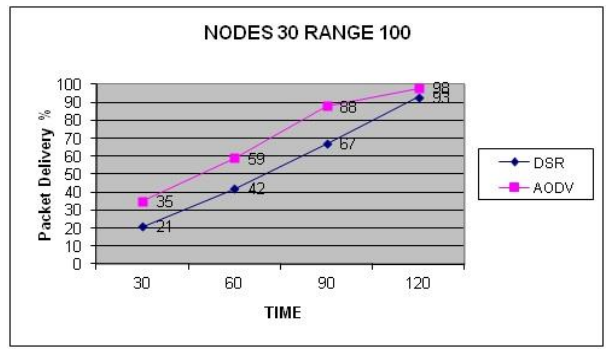

Fig 2: Packet Delivery Ratio in absence of malicious nodes.

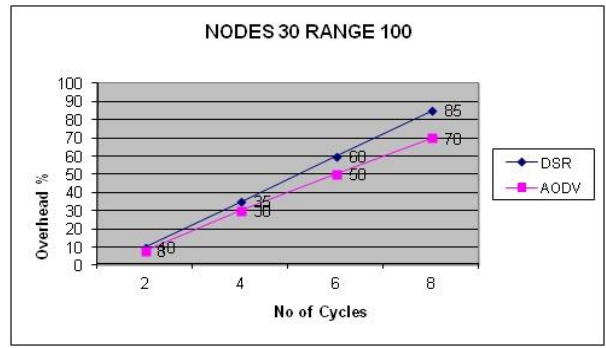

Fig 3 : Overhead ratio in the presence of malicious nodes

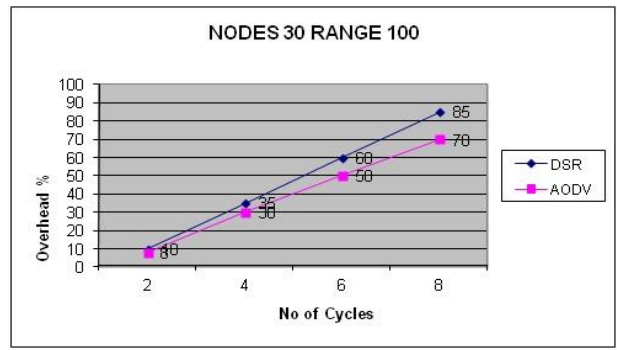

Fig 4 : Overhead ratio in the absence of malicious nodes

From the above fig 1 and fig 2, it is very clear that the packet delivery ratio is high in the presence of malicious nodes and is low in the absence of malicious nodes. When compared with the work done by other researchers where in their approach resulted in reduced packet loss in presence of malicious nodes in Trusted AODV (TAODV) protocol [18]. The results obtained here show more throughput in both DSR and AODV protocols with our proposed approach. And from fig 3 and fig 4, it is evident that the routing overhead is increases when there are no malicious nodes as all the malicious nodes are eliminated from the path if they are found to be misbehaving in DSR and AODV protocols. Therefore elimination of nodes results in less number of paths to reach the destination and therefore increases the overhead. In one of the similar kind of research, only DSR protocol was considered in eliminating the malicious nodes, which increases the overhead on the protocol in finding safe route to reach the destination [19]. Now in order to have high packet delivery ratio and to reduce the overhead, an approach of second chance is proposed here for each of the node based on trust and previous behavior. In other words instead of simply eliminating the node from the path, the reasons for its misbehavior could be studied and each node could be given a chance to get reintegrated into the network based on the reason how it misbehaved. The nodes after getting a second chance to get added in the network will prove to be fruitful as there will be more number of paths available resulting in high packet delivery ratio and reduced overhead as there will be less path rejections.

*if (node is malicious)*
*{*
*check the trust value and check past behavior;*
  *if (within tolerance)*
   *{*
    *forward data packet to next hop;*
    *update trust;*
   *}*
  *else*
   *{*
   *eliminate the node from the path;*
   *}*
 *else*
*{*
*forward data packet to best neighbor1;*
*}*
*}*

## CONCLUSION

The proposed approach intends to improve the packet delivery ratio and reduce the routing overhead caused in the absence of malicious node due to frequent topological changes in ad hoc networks. A second chance strategy for the reactive routing protocols like DSR and AODV is proposed to make a node reintegrated into the network considering the reasons for misbehavior of the nodes. A simulation analysis is done between DSR and AODV protocols both in the presence and absence of malicious nodes using the metrics packet delivery ratio and routing overhead to propose the second chance strategy for the node based on the trust rating. Using this strategy it will improve the packet delivery ratio as there will be more paths available to reach the destination and routing overhead will be reduced as there will be less path rejections. In future changes in these protocols can be done by using this approach and a simulative research can be done.